\documentclass[pra,twocolumn,superscriptaddress,longbibliography,
amsmath,amssymb,groupaddress,
]{revtex4-2}
\usepackage[colorlinks,allcolors=blue]{hyperref}

\usepackage{float}
\usepackage{physics,xcolor,palatino}
\usepackage{graphicx}
\usepackage{dcolumn}
\usepackage{bm}
\usepackage[normalem]{ulem}
\newcommand{\Id}{\mathbb{I}}
\newcommand{\ui}{\mathrm{i}}
\usepackage[T1]{fontenc}  

\makeatletter
\newcommand{\thickhline}{%
    \noalign {\ifnum 0=`}\fi \hrule height 1pt
    \futurelet \reserved@a \@xhline
}
\newcolumntype{"}{@{\hskip\tabcolsep\vrule width 1pt\hskip\tabcolsep}}
\makeatother

\begin{document}


 \title{Entanglement distribution through separable states via a zero-added-loss photon multiplexing inspired protocol}

\author{Conall J. Campbell}\thanks{These authors contributed equally to this work}
\affiliation{Centre for Quantum Materials and Technologies, School of Mathematics and Physics, Queen’s University Belfast, BT7 1NN Belfast, UK}

\author{Adam G. Hawkins}\thanks{These authors contributed equally to this work}
\affiliation{Centre for Quantum Materials and Technologies, School of Mathematics and Physics, Queen’s University Belfast, BT7 1NN Belfast, UK}

\author{Giorgio Zicari}
\affiliation{Centre for Quantum Materials and Technologies, School of Mathematics and Physics, Queen’s University Belfast, BT7 1NN Belfast, UK}

\author{Mauro Paternostro}
 \affiliation{Universit\`a degli Studi di Palermo, Dipartimento di Fisica e Chimica - Emilio Segr\`e, via Archirafi 36, I-90123 Palermo, Italy}
\affiliation{Centre for Quantum Materials and Technologies, School of Mathematics and Physics, Queen’s University Belfast, BT7 1NN Belfast, UK}

\author{Hannah McAleese}
\affiliation{Centre for Quantum Materials and Technologies, School of Mathematics and Physics, Queen’s University Belfast, BT7 1NN Belfast, UK}

\date{\today}

\begin{abstract}

{The recently proposed zero-added-loss multiplexing (ZALM) source of entangled photons enables higher efficiency in entanglement distribution than spontaneous parametric down-conversion sources and can be carried out using both space-to-ground and ground-to-ground links. We demonstrate the flexibility of ZALM architectures to be adapted to alternative entanglement distribution protocols. Focusing on the counterintuitive result that entanglement can be generated between distant parties without using any entanglement as a resource, we analyze two protocols for entanglement distribution to memories via separable states. Modeling them in a ZALM setup, we consider the effects of noise both in the communication channels and in the memories. We thereby identify  the optimal protocol to use with respect to the highest entanglement generated, given the noise conditions of the network.} 
\end{abstract}

\maketitle

\section{Introduction}

Quantum communication envisions a robust and scalable quantum internet, offering unprecedented capabilities in communications, notably with the potential for enhanced security through quantum key distribution~\cite{KimbleReview,WehnerReview}. The effective realization of the quantum internet heavily relies on the availability of suitable entangled resources, which should be shared among the elements of the communication network in a reliable and reproducible fashion~\cite{Belenchia2022}. 

However, direct entanglement distribution (DED), i.e. the unmediated sharing of entangled states among nodes of a quantum network, while intuitive, suffers from susceptibility to noise in the channel(s) between the nodes themselves. While remarkable steps forward have been made, recently, toward the achievement of reliable ways to distribute entanglement directly, including the pioneering demonstration of satellite-based DED~\cite{MiciusReview}, the establishment of long-haul entangled channels in quantum networks remains a practical challenge.

A  potentially fruitful way around the DED paradigm relies on the use of weaker forms of quantum correlations, particularly of the quantum discord type~\cite{Ollivier,Henderson}. In this context, approaches to the distribution of entanglement through states that, albeit separable, carry quantum discord have been put forward and demonstrated experimentally~\cite{Cubitt,Fedrizzi,Laneve,Chuan,Streltsov_Cost,McAleese,Vollmer,Peuntinger,Kay}. By leveraging separable states to catalyze the engineering of entangled remote nodes, such schemes capitalize on the robustness of quantum discord, which, unlike entanglement, might remain nonzero even under strong environmental influences, thus making it more resilient to certain types of noise. The corresponding schemes for entanglement distribution with separable states (EDSS) thus aim to overcome the challenges posed by noisy environments for quantum communication networks.

In this work, we compare the robustness of EDSS protocols with DED in the presence of noisy environments. All entanglement distribution protocols are simulated with a setup inspired by a physically motivated zero-added-loss photon multiplexing (ZALM) model~\cite{Chen}, which enables both space-to-ground and ground-to-ground communication. We also address the threat embodied by man-in-the-middle attacks and how ZALM-like schemes may be sabotaged with respect to the  quantum correlations established between Alice's and Bob's memories. We find that the ZALM architecture benefits from the intrinsic advantage stemming from EDSS protocols in the case of noiseless quantum memories. However, in the presence of  noise, such benefits are quickly lost, making DED a better option for distributing entanglement to spatially separated nodes of a network.

The remainder of this paper is structured as follows. In Sec.~\ref{entanglement_distribution_protocols sec}, we review two EDSS protocols and evaluate their performances -- in terms of the degree of entanglement achieved through them and their resilience to adversarial actions -- under ideal operating conditions. In Sec.~\ref{zalm mapping sec} we show that recently proposed ZALM architectures based on light-matter interaction could be exploited to achieve successful EDSS, thus making ZALM of more significant. Finally, in Sec.~\ref{simple_noise_models sec}, we discuss the performance of the EDSS protocols considered here under different noise models. Section~\ref{conc} offers concluding remarks and further perspectives on the topic of this study.

\section{Entanglement Distribution using Separable States}
\label{entanglement_distribution_protocols sec}
The idea of distributing entanglement to two spatially separated particles, which we label $A$ and $B$, using a carrier  that remains \textit{separable} from the other two at all times of the dynamics was introduced in Ref.~\cite{Cubitt}. The carrier particle, labeled $K$ herein, will remain separable in the $K:AB$ bipartition at every stage of the protocol. Following this early proposal, other entanglement distribution schemes based on the use of separable states have emerged for both discrete~\cite{Kay,Fedrizzi} and continuous-variable systems~\cite{Peuntinger}. The performance of such EDSS schemes can vary widely in terms of both the amount of distributed entanglement and the probabilities of success of the task. For instance, the scheme put forward by Cubitt \textit{et al.}~\cite{Cubitt} can result in a maximally entangled state with a success probability of $1/3$, while the distribution of entanglement using separable Bell-diagonal states pioneered by Kay~\cite{Kay} depends on both the initial states of particles $A$ and $B$ and the state of the carrier.

The resource able to unlock an EDSS approach is the availability of {\it weaker-than-entanglement} forms of quantum correlations, such as those quantified by {quantum discord}~\cite{Ollivier,Henderson,Modi2012}. That is to say, the amount of entanglement $\mathcal{E}_{\mathrm{dis}}$ that can be distributed to two parties through the scheme, intended here as the net difference between the entanglement ${\cal E}_{\rm initial}$ initially shared by the parties and the value ${\cal E}_{\rm final}$ finally achieved after application of the protocol, is bounded by the amount of \textit{communicated} discord $\mathcal{D}_{\mathrm{comm}}$ \cite{Fedrizzi,Chuan,Streltsov_Cost} as
\begin{equation}
    \mathcal{E}_{\mathrm{dis}} = \mathcal{E}_{\mathrm{final}} - \mathcal{E}_{\mathrm{initial}} \leq \mathcal{D}_{\mathrm{comm}}\,.
\end{equation}
Here, $\mathcal{D}_{\mathrm{comm}}$ is the degree of quantum discord established in the $K:AB$ bipartition.

We now briefly review two discrete-variable EDSS protocols, assessing the performance of each with the assumption of an ideal, noise-free setup.

\subsection{Protocol $\alpha$}
\label{Cubitt sec}
The first protocol we review stems from the example outlined in Ref.~\cite{Cubitt}, which considers the classically correlated initial state
\begin{equation}
\label{eq:rhoABK}
\begin{aligned}
    \rho _{ABK}&= \frac{1}{6}\left(\sum_{j=0}^3 \ketbra{\gamma_j,\gamma_{-j},0}{\gamma_j,\gamma_{-j},0}_{ABK}\right.\\ 
    &\left.+ \sum_{l=0}^1 \ketbra{l,l,1}{l,l,1}_{ABK}\right),
    \end{aligned}
\end{equation}
where $\ket{\gamma_j}_x = 1/\sqrt{2}(\ket{0}+e^{\ui j \pi/2}\ket{1})_x~(x=A,B)$. Tracing out $K$ from Eq.~\eqref{eq:rhoABK} leaves the $A-B$ compound in a separable state with discord $\mathcal{D}_{A|B}\simeq0.126$, as quantified by the difference in mutual entropy definitions \cite{Ollivier}. Alice initially holds her qubit $A$ along with the carrier qubit $K$ and performs a \textsc{cnot} operation between them, having $K$ as the target. We refer to this as  the encoding operation. She then sends $K$ to Bob, who subsequently performs the decoding operation on qubits $K$ and $B$. This consists of a \textsc{cnot} having $K$ as the target. The $A$-$B$-$K$ system is thus left in the state
\begin{equation}\label{Cubitt final state}
    \sigma_{ABK} = \frac{1}{3}\ketbra{\Phi^+}{\Phi^+}_{AB} \otimes \ketbra{0}{0}_K + \frac{2}{3}\Id_{AB}\otimes\ketbra{1}{1}_K\,,
\end{equation}
where $\ket{\Phi^+}_{AB}=1/\sqrt{2}(\ket{00}+\ket{11})_{AB}$ is the maximally entangled Bell state and $\Id$ is the identity operator. One can easily verify that the carrier $K$ remains separable from $AB$ at every stage of this protocol. The state after the encoding operation carries entanglement between $A$ and $B$-$K$: the decoding operation then localizes the $A:BK$ entanglement in the $A:B$ bipartition~\cite{Chuan}. By measuring $K$ in the computational basis with the condition of observing the state $\ket{0}$, we can achieve a Bell state with probability $1/3$. By using this postselective measurement or by entanglement distillation, as detailed in Ref.~\cite{Cubitt}, we can thus achieve maximum entanglement between $A$ and $B$, even though they do not interact directly with each other and despite the use of a carrier that is never entangled with them.

\subsection{Protocol $\beta$}
\label{Fedrizzi sec}
Reference~\cite{Kay} details a generalized EDSS protocol, in which Alice and Bob begin the protocol by sharing a separable Bell-diagonal state $\rho_{AB}$ and then use an initially {uncorrelated} carrier -- prepared in $\rho_K(c_x)=(\Id + c_x \sigma_x)/2$, where $\sigma_j$ is the Pauli $j$ matrix (with $j=x,y,z$) and $c_x\in [-1,1]$ -- to distribute entanglement. The experimental demonstration of this scheme reported in Ref.~\cite{Fedrizzi} had the combination of these states for the experiment such that the final distributed entanglement was maximized, which was obtained for $c_x=-1/2$ and the initial state
\begin{equation}
\label{fedrizzi state eq}
\begin{aligned}
    \rho_{AB} &= \frac{1}{4}\left(\sum_{j=0}^1 \ketbra{z_j z_j}{z_j z_j}_{AB} + \frac{1}{2} \sum_{j=0}^1 \ketbra{x_j x_j}{x_j x_j}_{AB}\right.\\
    &\left.+ \frac{1}{2} \sum_{j=0}^1 \ketbra{y_j y_{1-j}}{y_j y_{1-j}}_{AB}\right),
\end{aligned}
\end{equation}
where $\ket{k_j}$ is the eigenstate of the Pauli $k$ matrix with eigenvalue $(-1)^j$. Despite being separable due to its invariance under partial transposition, this state is endowed with a degree of discord of $\mathcal{D}_{A|B}\simeq 0.0613$.

\begin{figure*}[ht]
 {\bf (a)}\hskip6cm{\bf (b)}\\
         \includegraphics[width=0.95\columnwidth]{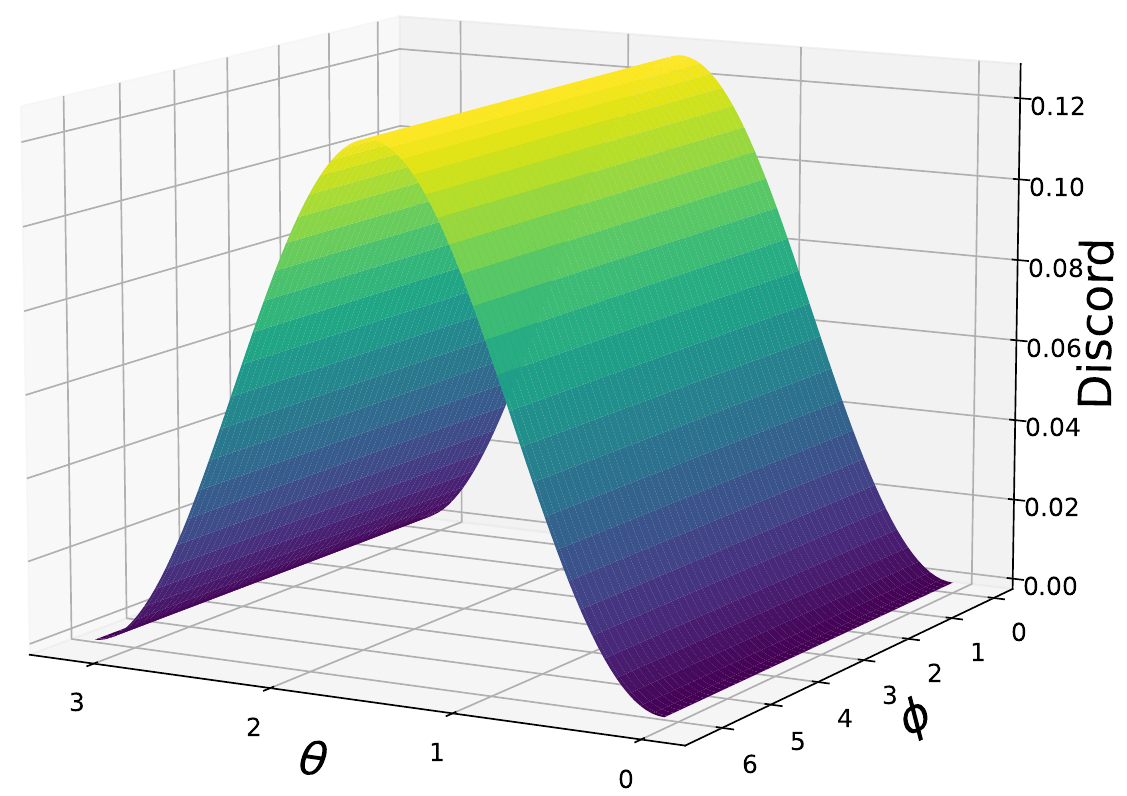}
         \includegraphics[width=0.95\columnwidth]{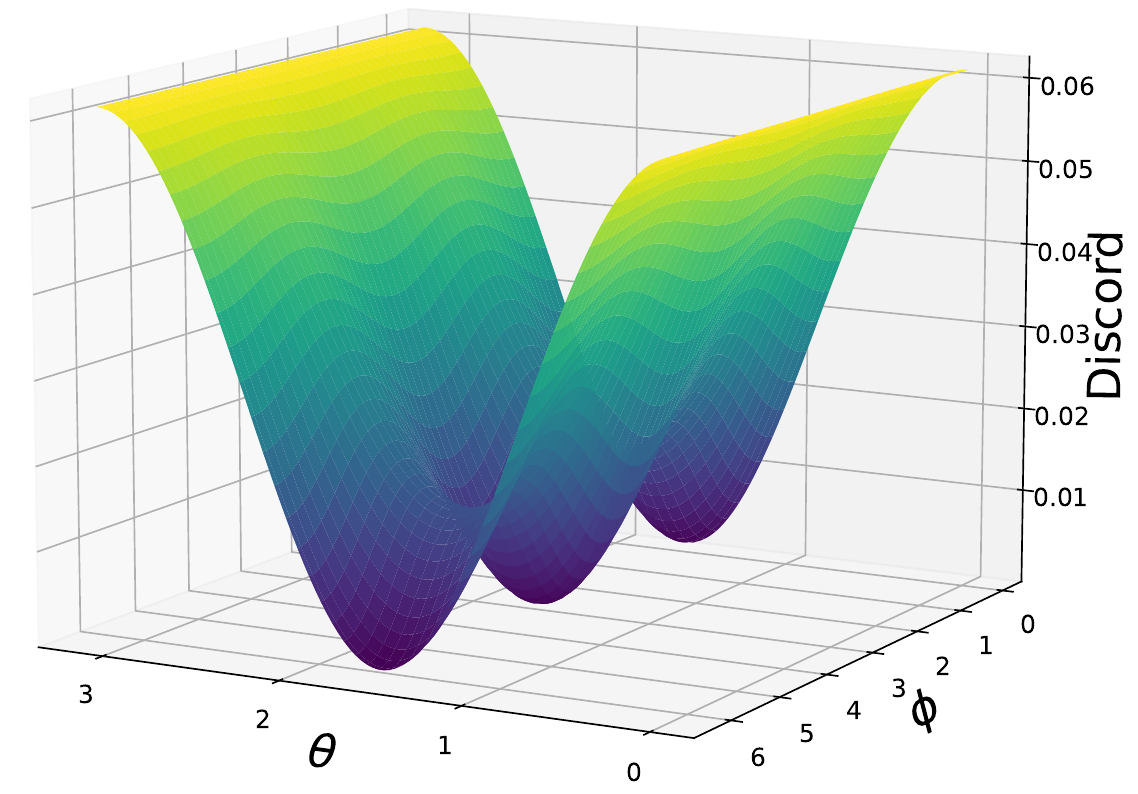}
     \caption{Degree of quantum discord in the $A$-$B$ state upon measurement of $K$ by an adversarial agent Eve, plotted against the Bloch parameters $\theta$ and $\phi$. The  measurement occurs after the encoding operation between $A$ and $K$. Trend corresponding to {\bf (a)} protocol $\alpha$ and {\bf (b)} protocol $\beta$.}
     \label{adversary plots}
\end{figure*}

The protocol then proceeds as follows: Alice performs a controlled-phase (\textsc{cphase}) encoding  operation on qubits $A$ and $K$ and then sends $K$ to Bob, who performs a   \textsc{cphase} gate on $K$ and $B$ as a decoding transformation. After the encoding step, the $A:BK$ bipartition is entangled, as seen by using the negativity 
\begin{equation}
\label{negativity_eq}
    \mathcal{N}_{A:BK} = \sum_l \frac{\abs{\lambda_l}-\lambda_l}{2}
\end{equation}
to quantify it~\cite{Peres,Horodecki}. Here, $\{\lambda_l\}$ is the set of eigenvalues of the partial transposition of Eq.~\eqref{fedrizzi state eq} with respect to either $A$ or the compound $B$-$K$. We find $\mathcal{N}_{A:BK}=1/16$. As in protocol $\alpha$, such entanglement can be localized to $A:B$ via the decoding operation, so that a postselective local measurement on $K$ can be performed. Such localized entanglement reaches a maximum value of $\mathcal{N}_{A:B}=1/10=0.1$ when $K$ is measured in the $\{\ket{x_0},\ket{x_1}\}$ basis and the state $\ket{x_1}$ is observed \cite{McAleese}, which occurs with probability $5/8$. It should be noticed that a maximally entangled pure state of two qubits attains a value of negativity of 0.5 according to Eq.~\eqref{negativity_eq}.

Although the resulting entanglement is distillable, it can also be raised via repeated iteration of the 
protocol based on the use of a stream of separable carriers $K_i$, each prepared in state $\rho_{K_i}(-1/2)$ for the $i$th iteration. After the respective encoding operation at Alice's site, $K_i$ is sent to Bob for decoding and postselective measurement in the $\{\ket{x_0},\ket{x_1}\}$ basis. Upon the first successful iteration of this protocol, we achieve ${\cal N}_{A:B}\simeq0.143$. Further iterations of such a process result in ${\cal N}_{A:B}\to1/6$ as $i\to\infty$, with each carrier remaining separable at all times in all of the iterations that we have considered (see Sec.~\ref{simple_noise_models sec}).

Finally, it is worth noting that, while in this work we focus on the bipartite case, the protocol outlined here can be developed and generalized to networks with more than two nodes~\cite{Laneve}.

\subsection{Effect of an adversarial agent}

We now consider the action of an adversarial agent, whom we call Eve, attempting to sabotage the scheme by making a measurement on the carrier $K$, following the encoding operation. We model such measurement using the projector $\Pi_K=\ketbra{\psi}{\psi}_K$, with $\ket{\psi}_K=\cos(\theta/2)\ket{0}_K+\sin(\theta/2)e^{\ui\phi}\ket{1}_K$ defined by the Bloch parameters $\theta\in[0,\pi]$ and $\phi\in[0,2\pi)$. The postmeasurement discord between $A$ and $B$ for both protocols is plotted as a function of such parameters in Fig.~\ref{adversary plots}.

For protocol $\alpha$, a measurement by the adversarial agent in a basis where $\theta=\pi/2$ leaves $A$ and $B$ with the same degree of shared discord with which they started the protocol. Examples of such bases include $\{\ket{x_0},\ket{x_1}\}$ and $\{\ket{y_0},\ket{y_1}\}$. For protocol $\beta$, the amount of discord in the initial state is recovered if Eve measures in $\{\ket{z_0},\ket{z_1}\}$, with $\ket{z_0}$ leaving $A$ and $B$ in the state they were initially in before the encoding operation. In this case, repetition of the protocol will be straightforward, as $K$ is initially completely uncorrelated with the state of the $A$-$B$ compound. A measurement in $\{\ket{x_0},\ket{x_1}\}$ destroys all discord, while a measurement in $\{\ket{y_0},\ket{y_1}\}$ leaves them in a state with reduced discord.

Eve could also sabotage the protocols by applying a local  rotation to $K$ after the encoding operation involving $A$. While this  would not affect the amount of discord in the state~\cite{Modi2012}, the optimal basis for the scope of achieving a large amount of entanglement in the $A:B$ bipartition that Bob should use to measure $K$ would depend on the rotation itself. Eve could thus act in a way to deplete the entanglement  between qubits $A$ and $B$, all the way to separability.

\section{ZALM-based EDSS setting}\label{zalm mapping sec}

The principle of entanglement distribution underpinned by EDSS is suited to adaptation to architectures for entanglement-based quantum networking \cite{Murao,Dur,Karlsson,Zhao,Hillery,Cleve,Fitzi,Cabello} and distributed quantum computation~\cite{Caleffi2024}. An interesting example, suitable for both space-to-ground and ground-to-ground arrangements, has been put forward through the ZALM scheme of Ref.~\cite{Chen}. In this proposal,  a pair of spontaneous parametric down-conversion sources is used to produce heralded Bell pairs \cite{Dhara} in such a way that temporal-spectral multiplexing can be achieved from the large phase-matching bandwidth. The proposed scheme eliminates switching losses due to multiplexing in the source, and it is expected to facilitate the \textit{downlink} \cite{Pirandola} transmission for space-to-ground communication. 

While the scheme was designed to distribute entanglement directly to remote nodes of a network, here, we bring to the fore a different version of ZALM that leverages EDSS in an attempt to bypass the fragility of entanglement communicated across long-haul channels. For instance, the distributed entanglement is severely affected by dephasing noise, which might even induce its finite-time disappearance~\cite{Yu,YuReview}, contrary to the behavior of quantum discord, which, under the action of such a channel, exhibits only asymptotic decay \cite{Werlang}. We thus propose a different paradigm in which separable discordant photonic states are sent to the remote nodes of a network onto whose state they are then mapped. Concretely, in line with the proposal in Refs.~\cite{Chen,ChenNano}, one can consider the electronic-spin degree of freedom of a color center in diamond hosted in an optical cavity to implement the memories at each node, thus taking advantage of the long optical-spin coherence time of such a system~\cite{ChenNano}. The photon-to-spin mapping mechanism would then follow the path illustrated in Ref.~\cite{Chen}, to which we refer. 

The protocol for a general state of two qubits proceeds as follows. Consider a general mixed state $\rho_{P,\mathrm{ini}}$ of two photonic qubits $P=\{P_A,P_B\}$ encoded in the horizontal and vertical polarization states $\{\ket{H},\ket{V}\}$. The spatially separated matterlike memories that embody the nodes of the network we are interested in are encoded in the electron-spin states $\{\ket{\downarrow},\ket{\uparrow}\}$ of qubits $S=\{A,B\}$. The spin system $S$ is initially prepared in $\rho_{S,\mathrm{ini}}=\ketbra{x_0x_0}{x_0x_0}_{AB}$. 

The photon state would thus enter a polarization splitter used to convert the polarization basis into the spatial one $\{\ket{a_H},\ket{a_V}\}$ with $a_{H,V}$ distinguishable spatial modes, determined by the states of the polarization basis and addressing them to the respective cavities. Mode $a_H$ would acquire a spin-dependent phase upon cavity reflection (if the spin is in state $\ket{\uparrow}$), whereas mode $a_V$ is a constant $-1$ phase regardless of the spin state \cite{Chen,DuanPhotonSpin,DuanAtomGates}. Writing this in the form of an operator, we have the two-qubit gates $U_{P_JJ} = \ketbra{a_H}{a_H}_{P_J}\otimes\ketbra{\uparrow}{\uparrow}_{J} - \ketbra{a_H}{a_H}_{P_J}\otimes\ketbra{\downarrow}{\downarrow}_{J} - \ketbra{a_V}{a_V}_{P_J}\otimes \Id_{J}$
for subsystem $P_JJ$, with $J=\{A,B\}$. After this interaction, the photonic states enter a 50:50 beam splitter with output modes $\ket{\mathcal{A}_\pm}_P = (\ket{a_H} \pm \ui\ket{a_V})/{\sqrt{2}}$. The photons are measured in the $\{\ket{\mathcal{A}_\pm}_P\}$ basis at each respective node. We model these measurements using the rank-1 projectors $\Pi^{\mathcal{A}_\pm}_P = \ketbra{\mathcal{A}_\pm}{\mathcal{A}_\pm}_P = \frac{1}{2}(\Id \pm \sigma_y)$. There are thus four different possible states for the spins to be in after the measurement, depending on which of the four measurement outcome combinations for qubits $P_A$ and $P_B$ is observed. Such normalized states are given by
\begin{equation}
    \rho_{\mathrm{out}}^{jk} =\frac{(\Pi_{P_A}^j\otimes\Pi_{P_B}^k) \rho_{BS} (\Pi_{P_A}^j\otimes\Pi_{P_B}^k)}{\Tr{(\Pi_{P_A}^j\otimes\Pi_{P_B}^k) \rho_{BS} (\Pi_{P_A}^j\otimes\Pi_{P_B}^k)}}\,,
\end{equation}
where the indices $j,k=\{\mathcal{A}_\pm\}$ correspond to the measurement outcome $\ket{jk}_{P_AP_B}$ of the photonic qubits and $\rho_{BS}$ is the state of $P_AP_B AB$ after the photons pass through the beam splitter. Each of the four resulting states occurs with equal probability $p^{jk}={1}/{4}$.

\begin{figure}[t!]
    \centering
    \includegraphics[width=\linewidth]{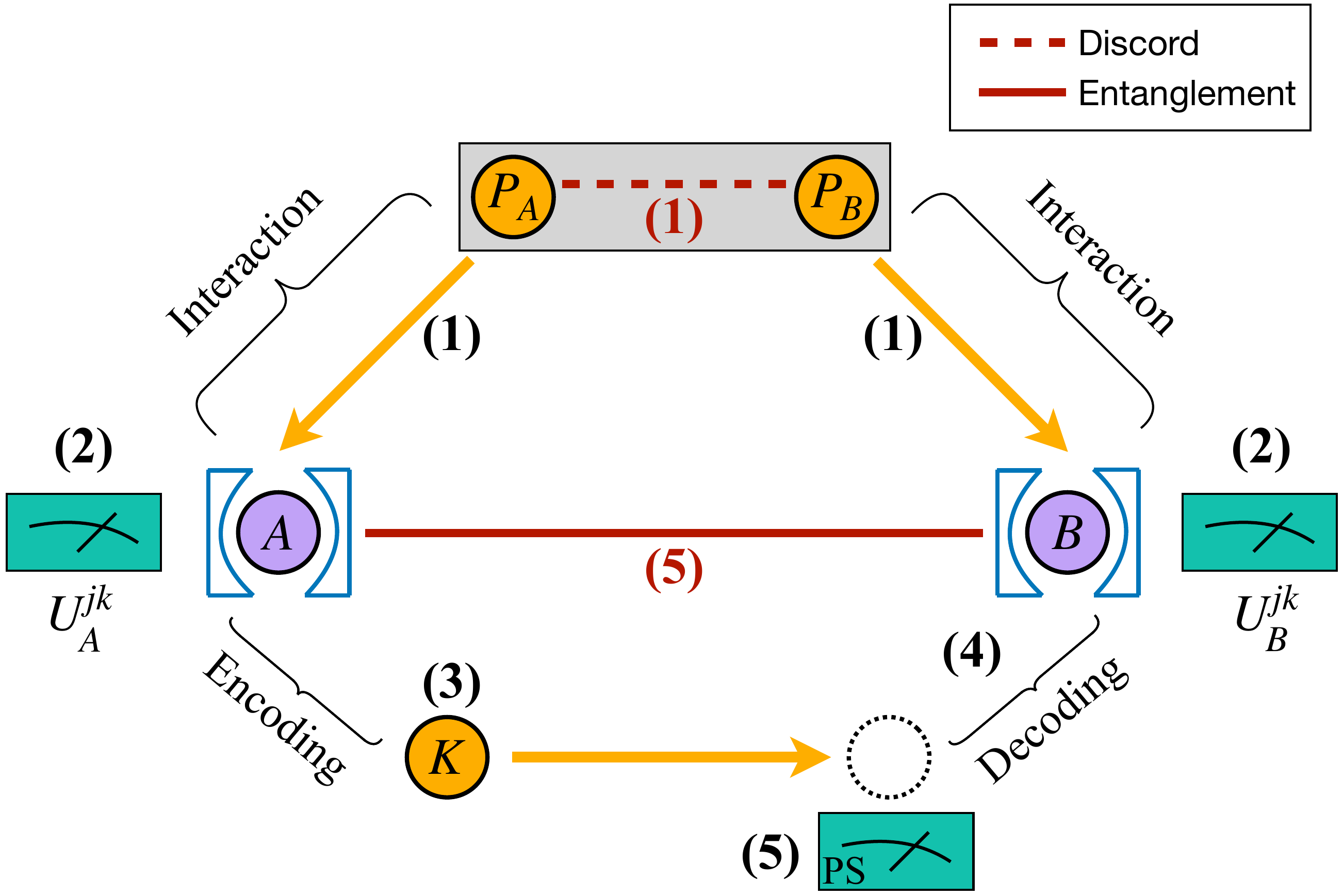}
    \caption{An EDSS protocol utilizing a ZALM-type architecture. \textbf{(1)} A pair of photons $P_{A,B}$ is prepared in a discordant, but separable, state. Photon $P_A$ ($P_B$) is sent to memory $A$ ($B$) for a \textsc{cphase} interaction. Memories take the form of atoms suspended in optical cavities. \textbf{(2)} Each photon passes through a 50:50 beam splitter and is then measured. A measurement-dependent unitary operation ${U_A^{jk}}$ ($U_B^{jk}$) is applied to memory $A$ ($B$) as described in Table.~\ref{tab: measurement dependent operations}). \textbf{(3)} Alice introduces a separable carrier photon $K$ and performs an encoding operation. \textbf{(4)} $K$ is sent to Bob, who performs the decoding operation. \textbf{(5)} A postselective (PS) measurement is performed on $K$, which remains separable from $AB$ at all times. $A$ and $B$ now share entanglement.}
    \label{fig: protocol diagram}
\end{figure}

We then trace out $P_A$ and $P_B$, and apply Hadamard gates $H$ to qubits $A$ and $B$, leaving the spin qubits in the state
\begin{equation}
    \rho_{S,\mathrm{out}}^{ij} =(H_A\otimes H_B)\Tr_{P_AP_B}\left\{\rho_{\mathrm{out}}^{jk}\right\}(H_A\otimes H_B)\,.
\end{equation}
Finally, a local operation $U_S^{j,k}$ must be applied to each spin qubit, depending on the measurement outcome of the respective photonic qubit. Each operation $U_S^{j,k}$ can be decomposed as 
\begin{equation}\label{eq: local unitaries after measurement}
    U_S^{jk} = U_A^{j} \otimes U_B^{k} = \begin{pmatrix}
        0&1\\
        e^{\ui\varphi_j}&0
    \end{pmatrix}
    \otimes \begin{pmatrix}
        0&1\\
        e^{\ui\varphi_k}&0
    \end{pmatrix}\,,
\end{equation}
with $\varphi_{\mathcal{A}_\pm} = \mp\pi/2$. This is summarized in Table~\ref{tab: measurement dependent operations}. The final spin state $\rho_{S,\mathrm{fin}} = U_S^{jk}\rho_{S,\mathrm{out}}^{jk}U_S^{jk^\dagger}$ can be shown to be equivalent to the initial state of the photonic qubits irrespective of the initial state of the $P$ compound, i.e., $\rho_{S,\mathrm{fin}} \equiv \rho_{P,\mathrm{ini}}$.

\begin{table}
    \centering
    \begin{tabular}{c|c"c|c}
         \multicolumn{2}{c"}{BS outcome} & \multicolumn{2}{c}{Phase Angle}\\
         $\,  \ket{j}_{P_A}$ \  & \,  $\ket{k}_{P_B}$ \  & \, \ $\varphi_j$ \, \,  & $\varphi_k$ \\
         \thickhline
        $\ket{\mathcal{A}_+}$ & $\ket{\mathcal{A}_+}$ & $-\pi/2$ & $-\pi/2$ \\
         \hline
        $\ket{\mathcal{A}_+}$ & $\ket{\mathcal{A}_-}$ & $-\pi/2$ & $+\pi/2$ \\
         \hline
         $\ket{\mathcal{A}_-}$ & $\ket{\mathcal{A}_+}$ & $+\pi/2$ & $-\pi/2$ \\
         \hline
         $\ket{\mathcal{A}_-}$ & $\ket{\mathcal{A}_-}$ & $+\pi/2$ & $+\pi/2$ \\
    \end{tabular}
    \caption{The appropriate phase angles $\varphi_{j,k}$ that define the local unitary operations in Eq.~\eqref{eq: local unitaries after measurement} which are applied to the spin qubits. These angles depend on which of the beam splitter (BS) output modes is observed for each photon qubit.}
    \label{tab: measurement dependent operations}
\end{table}

Using this approach, a discordant state can be mapped onto the remote qubit memories. If the mapping is successful, we can then perform an EDSS protocol using the discordant state of the memories as a resource and introducing a third, separable, carrier photon $K$. This may be realized by again performing \textsc{cphase} encoding and decoding operations implemented via dispersive light-matter interactions involving the cavity-embedded electron-spin systems and a photon sent from $A$ to $B$~\cite{ChenNano}. The entire protocol is illustrated in Fig. \ref{fig: protocol diagram}. 

\section{Simulating the performance of entanglement distribution schemes under simple noise models}
\label{simple_noise_models sec}

An aspect that should be carefully considered when addressing  the realization of long-distance EDSS protocols, particularly when evaluating their effectiveness against alternatives based on direct entanglement distribution approaches, is the impact of noise in the signal from the discordant source to the memories, as well as between the memories themselves. Noise could severely affect the performance of the scheme, all the way to the full loss of any quantum resource. 

In order to quantitatively address such issues and determine the effectiveness of the ZALM architecture for EDSS against DED approaches, we subject both protocols discussed in Sec.~\ref{entanglement_distribution_protocols sec} and DED to different noise models. We then build a comparative assessment with a ZALM-like DED scheme in which a source in the middle 
generates a Bell pair $\ketbra{\Phi^+}{\Phi^+}$, whose qubits are then distributed to Alice and Bob. The performance of the protocols is assessed by comparing the degree of negativity [see Eq.~\eqref{negativity_eq}] localized in the states of the memories.

In Sec.~\ref{single channel model sec}, we begin by addressing the performance of the key resource for each ED scheme in noisy environments, which we model with depolarizing, dephasing, or amplitude damping noise. For DED we assume that the external third party sends the entangled qubits to Alice and Bob through noisy quantum channels. This is modeled as
\begin{equation}
\label{DED_noise eq1}
        \rho_{AB} = \sum_{i,j}\varepsilon^{ij}_{AB}(p) \; |\Phi^{+}\rangle_{AB}\langle\Phi^{+}| \; \varepsilon^{ij\dag}_{AB}(p)
        \end{equation}
{for } $\varepsilon^{ij}_{AB}(p) = M_{i , A}(p) \otimes M_{j , B}(p)$. Here, $M_{j,X}(p)$ denotes the $j$th Kraus operator of a particular noisy quantum channel applied on qubit $X$ with probability {$p\in[0,1]$}. For the EDSS schemes, we consider sending only the separable carrier $K$ through a noisy quantum channel from Alice's site to Bob's before we perform the decoding operation. This is modeled as
\begin{equation}
  \label{general_simple_channel_applied}
\rho'_{ABK}=     \sum_{j} \big[\mathbb{I}_{AB} \otimes M_{j,K}(p) \big] \rho_{ABK} \big[\mathbb{I}_{AB} \otimes M^{\dagger}_{j,K}(p) \big]
\end{equation}
As discussed in Sec.~\ref{Fedrizzi sec}, multiple iterations of protocol $\beta$ increase the entanglement localized between the cavities, and the carrier remains separable throughout these repetitions. In order to gauge the performance of the protocol against noise, we considered up to four iterations, which was the largest number that could be considered in our simulations without unnecessary computational burden while providing quantitatively and qualitatively informative results.

In Sec.~\ref{multi channel model sec}, we consider a simulation of the EDSS schemes that takes the noise of the memories into account before we perform the encoding operation and also considers the effect of a channel on the carrier before the decoding operation. This results in the following model:
 \begin{equation}
 \label{multi channel eq 1}
         \rho_{ABK}'' = \sum_{k} \big[ \mathbb{I}_{AB} \otimes M_{k , K}(p) \big] \tilde{\rho}_{ABK} \big[ \mathbb{I}_{AB} \otimes M_{k , K}^\dagger(p) \big]
\end{equation}
with $\tilde{\rho}_{ABK}$ being the state resulting from the effect of noise on the $A$-$B$ compound, given by
\begin{equation}
\label{multi channel eq 2}
    \begin{split}
        \tilde{\rho}_{ABK} &= \sum_{i,j}\varepsilon'^{ij}_{ABK}(p)\, \rho_{ABK} \,\varepsilon'^{ij\dag}_{ABK}(p)\\
        \varepsilon'^{ij}_{ABK}(p) &=  M_{i , A}(p)\, \otimes\, M_{j , B}(p)\, \otimes\, \mathbb{I}_K.
    \end{split}
\end{equation}

In this work, we optimize over all possible projective measurements to obtain the optimal measurement outcome that maximizes $AB$ entanglement, and determine how the amount of localized entanglement changes with noise.

\subsection{Single-channel model}
\label{single channel model sec}

\begin{figure}[ht!]
{\bf (a)}\\
\includegraphics[width=\columnwidth]{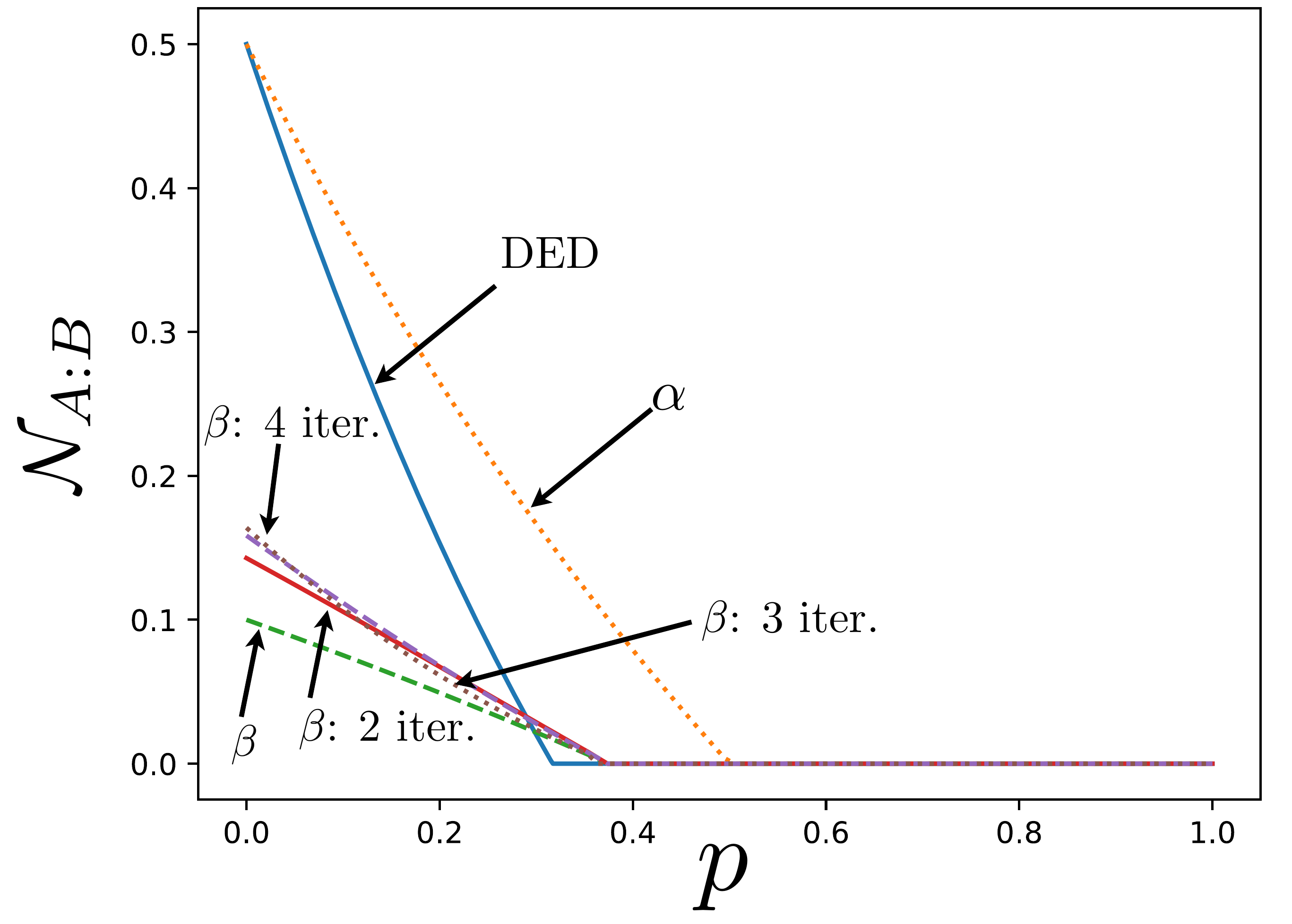}\\
{\bf (b)}\\\includegraphics[width=\columnwidth]{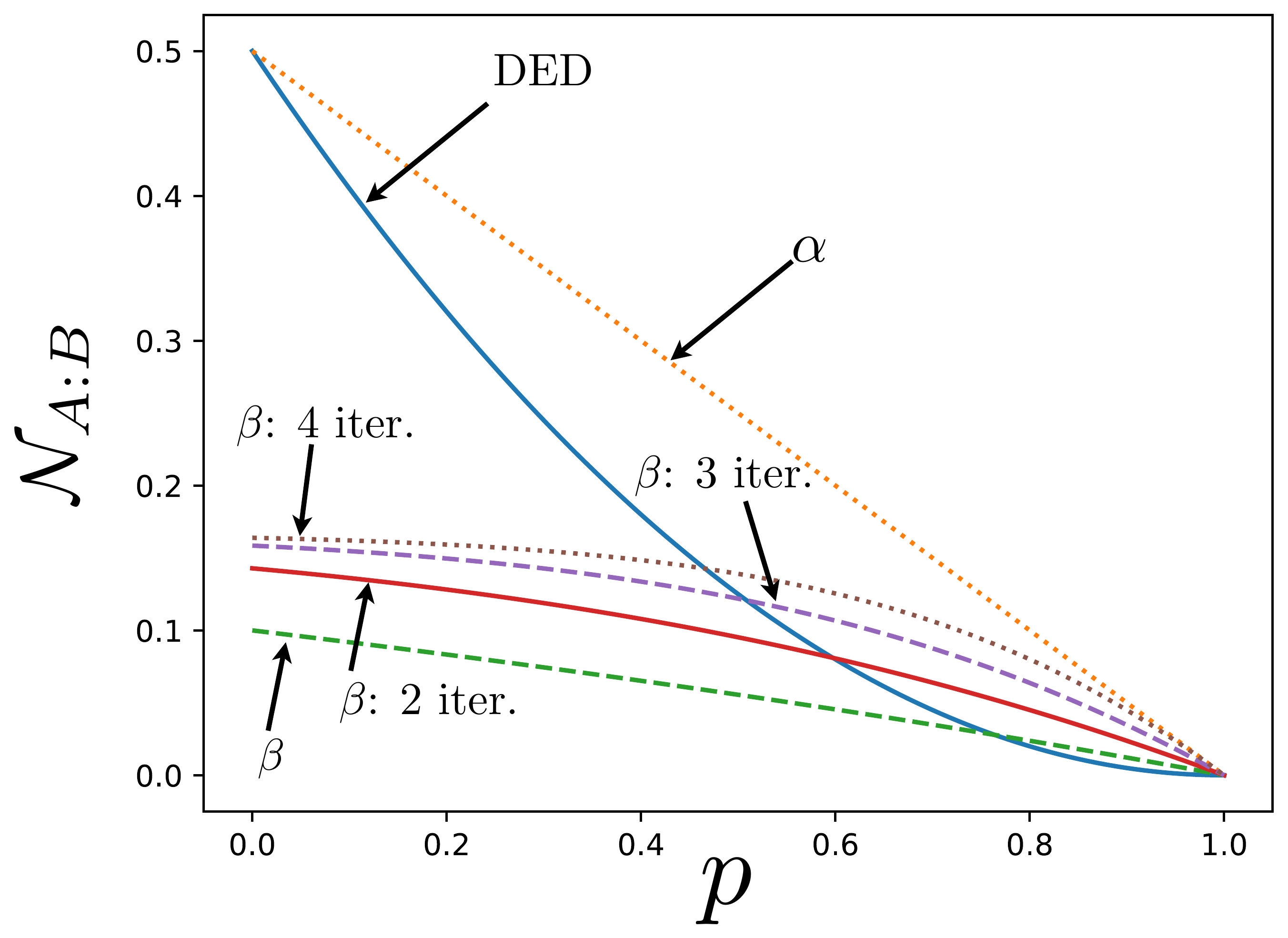}\\
{\bf (c)}\\
\includegraphics[width=\columnwidth]{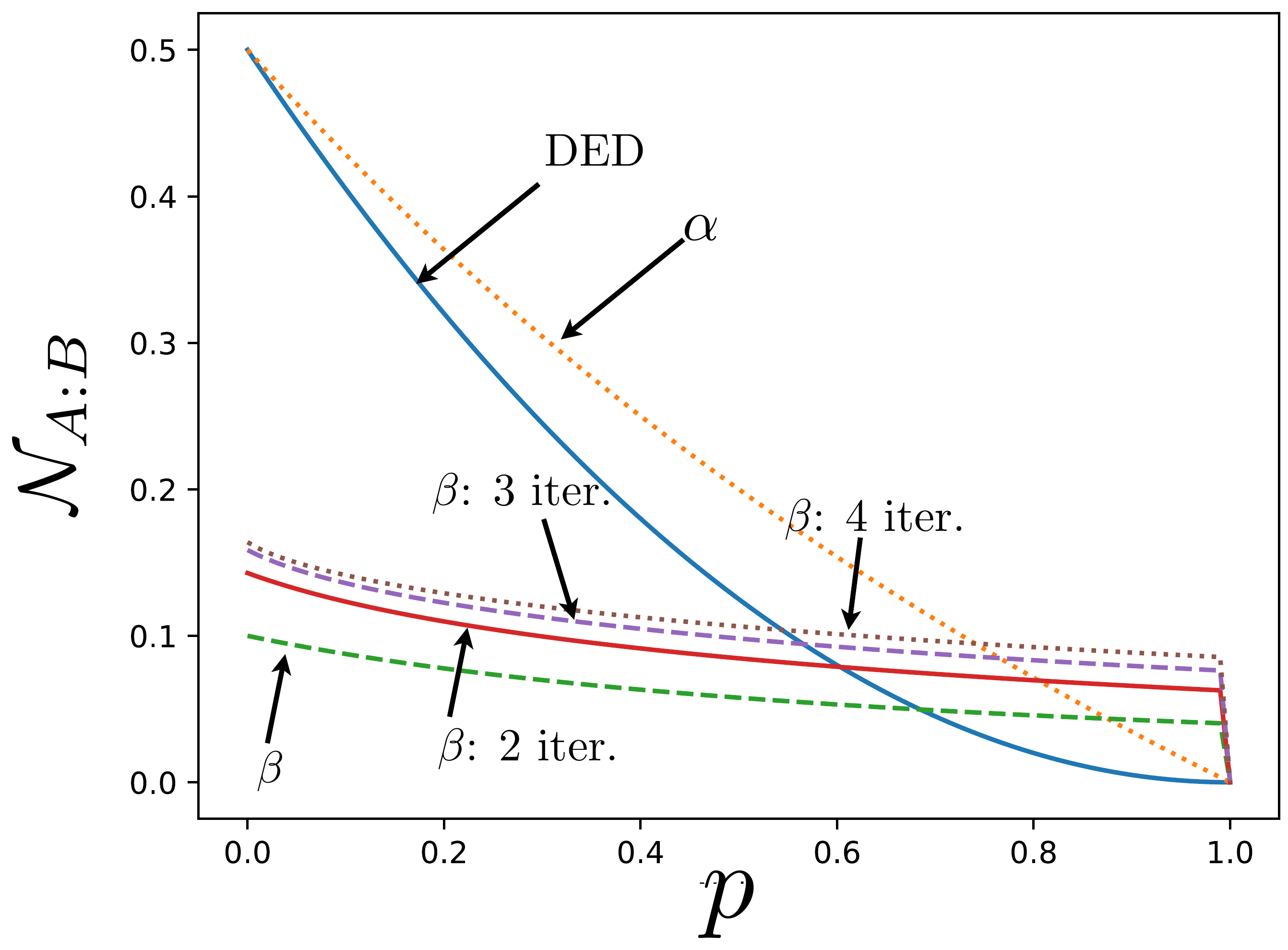}
     \caption{Behavior of the entanglement shared between $A$ and $B$ as a result of the three protocols considered in Sec.~\ref{single channel model sec}. We plot the negativity ${\cal N}_{A:B}$ against the strength $p$ of {\bf (a)} a depolarizing channel, {\bf (b)} a dephasing channel, and {\bf (c)} an amplitude damping one. Protocol $\beta$ is iterated up to four times.}
     \label{fig:channels_figure}
\end{figure}

From Eqs.~\eqref{DED_noise eq1} and~\eqref{general_simple_channel_applied}, we test the performance of the ED schemes when the key resource of the protocols is subjected to any of the aforementioned noise models. First, we consider the depolarizing channel; for a probability $(1-3p/4)$ the qubit remains unspoiled, and for a probability $p/4$ the qubit can be subjected to a phase-flip error, a bit-flip one, or both~\cite{Preskill}. The channel is represented in operator-sum form by the Kraus operators 
\begin{equation}
    \label{depolarising_channel_kraus}
        M_{1}(p)=\sqrt{\frac{p}{3}}\sigma_x,
         M_{2}(p)=\sqrt{\frac{p}{3}}\sigma_y,\, M_{3}(p)=\sqrt{\frac{p}{3}}\sigma_z,
\end{equation}
with $p\in[0,1]$ and $M_0(p)$ deduced from the completeness relation $\sum^3_{j=0}M^\dag_j(p)M_j(p)=\mathbb{I}$. The behavior of the three protocols with respect to the strength of the noise $p$ is investigated in Fig.~\ref{fig:channels_figure} {\bf (a)}, where we see that, while a sudden disappearance of entanglement occurs for all protocols, the EDSS ones are more advantageous than DED, as they show more robustness.

Next, we consider the action of a dephasing channel which models the decoherence of a qubit, transforming it into a mixed state with no off-diagonal elements with probability $p$ for a given measurement basis. The channel is represented by the Kraus operators
\begin{equation}
    \label{dephasing_channel_kraus}
        M_{0}(p){=}\sqrt{1-p}\mathbb{I},\,M_{1}(p){=}\sqrt{p}\begin{pmatrix}1&0\\
        0&0\\ \end{pmatrix},\,M_{2}(p){=}\sqrt{p}\begin{pmatrix} 0&0\\     0&1\\ \end{pmatrix}.
\end{equation}
Figure~\ref{fig:channels_figure} {\bf (b)} demonstrates the resilience of the EDSS protocols under the action of a dephasing channel. One can see that all discord-based protocols not only outperform direct ED but can localize entanglement between the memories in regions of high noise, thus demonstrating the importance of the  robustness of discord.

Last, we consider an amplitude damping channel, which can model photon loss or the decay of an excited two-level atom. The channel is represented by the Kraus operators
\begin{equation}
    \label{amplitude_damping_channel eq}
    M_{0}(p)=
    \begin{pmatrix}
    1&0\\
    0&\sqrt{1-p}\\
    \end{pmatrix}\text{ },\text{ }
    M_{1}(p)=
    \begin{pmatrix}
    0&\sqrt{p}\\
    0&0\\
    \end{pmatrix}.
\end{equation}
Figure~\ref{fig:channels_figure} {\bf (c)} once again illustrates the benefit of using quantum discord as a resource in ED protocols. Although protocol $\alpha$ offers a significant advantage over DED in this framework, single and multiple iterations of protocol $\beta$ are particularly beneficial because they can localize entanglement in highly noise affected conditions, once again highlighting discord's remarkable durability in the presence of noise.

The analysis in this section shows that, while the consideration of multiple iterations of protocol $\beta$ is, indeed, generally successful in localizing a larger value of entanglement in the state of the memories, the largest gain is achieved in going from the first  iteration to the second: Additional iterations have only an incremental effect on the performance of just two iterations.

The probability of localizing maximum entanglement between memories $A$ and $B$ at the end of protocols $\alpha$ and $\beta$ depends on the strength of the noise $p$ exhibited by the quantum channel through which the separable carrier $K$ is transmitted and also the type of noise experienced by $K$ within the channel.
\begin{figure*} 
{\bf (a)}\hskip8cm{\bf (b)}\\
    \includegraphics[width=\columnwidth]{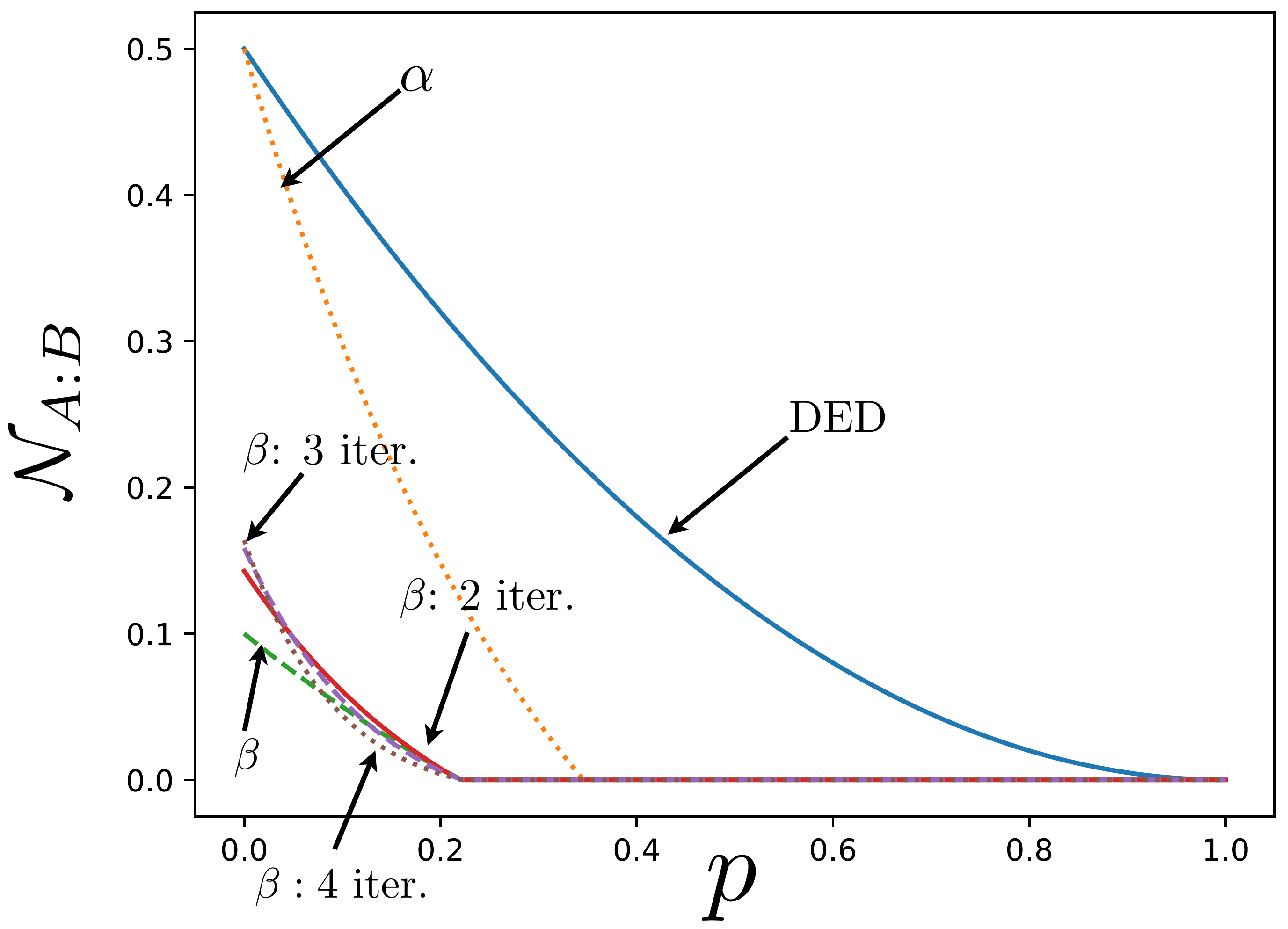}
    \includegraphics[width=\columnwidth]{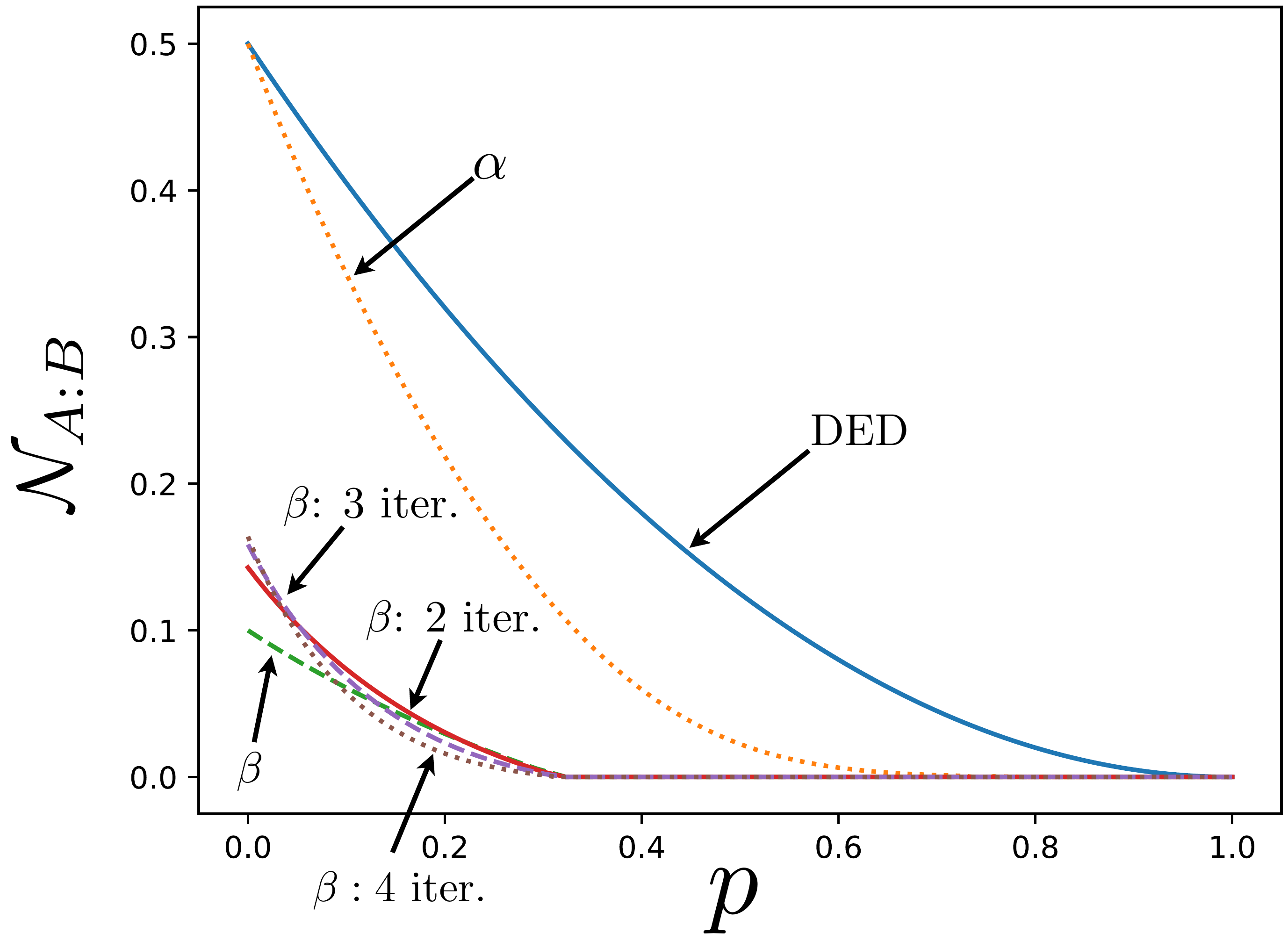}\\
    {\bf (c)}\hskip8cm{\bf (d)}\\
    \includegraphics[width=\columnwidth]{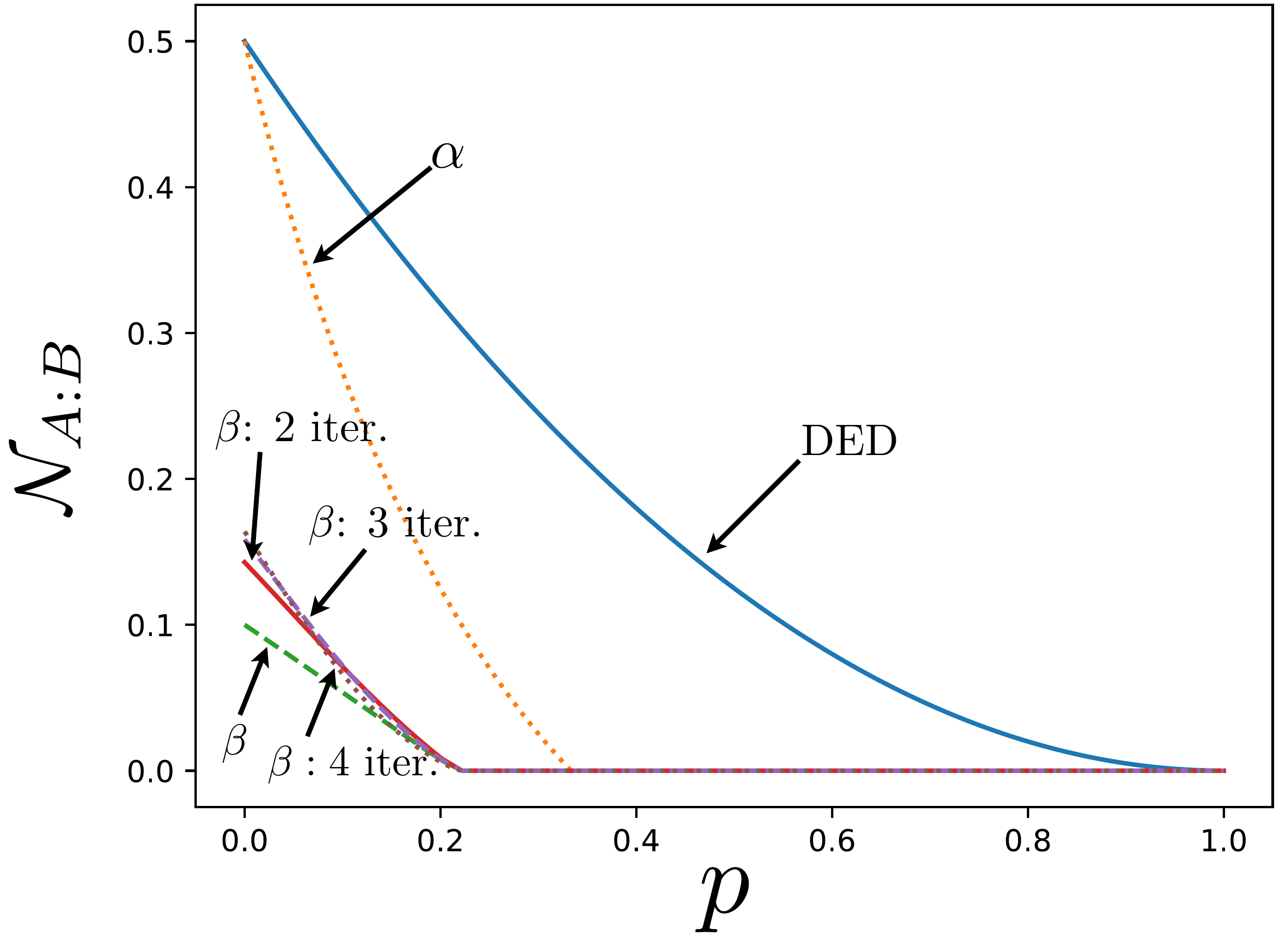}
    \includegraphics[width=\columnwidth]{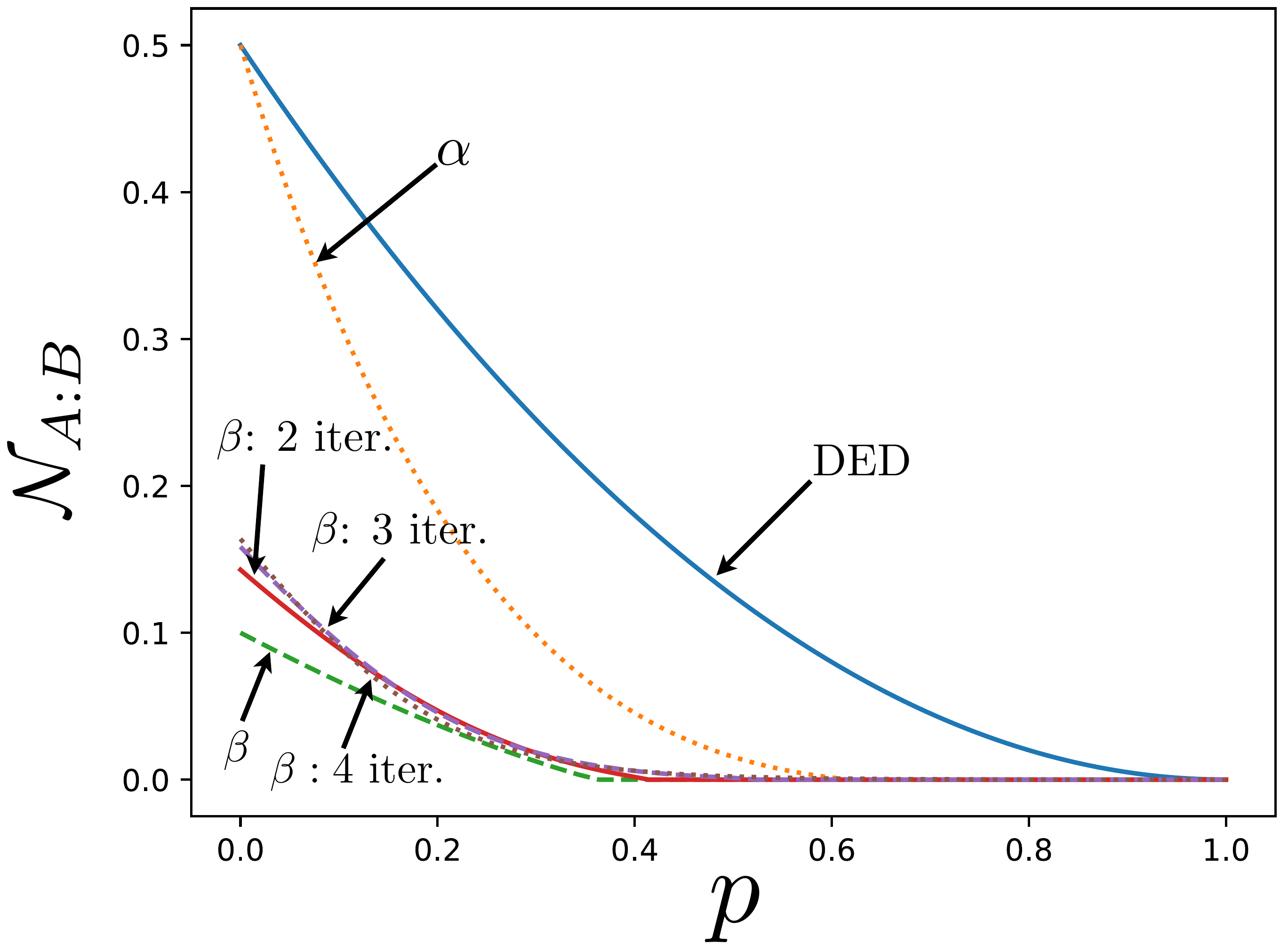}
    \caption{Plots showing the performance of the DED and EDSS protocols under the effects of channels with equal strengths. The top row considers the case when the quantum memories are exposed to dephasing noise and the separable carrier is exposed to \textbf{a} depolarizing noise and \textbf{b} amplitude damping noise. The bottom row considers the case where the quantum memories are exposed to amplitude damping noise and the separable carrier is exposed to \textbf{c} depolarizing noise and \textbf{d} amplitude damping noise.}
    \label{fig: same_strength_channels plots}
\end{figure*}

In the case of the aforementioned noise models, the probability of obtaining the right measurement outcome for protocols $\alpha$ and $\beta$ varies with $p$ as
\begin{equation}
\begin{aligned}
    \label{probability_eq}
    p_{\alpha , \mathrm{depo}} = \frac{1}{3} + \frac{2}{9}p \text{ } , \text{ } p_{\beta,\mathrm{depo}} &= \frac{5}{8} - \frac{1}{6}p \\
    p_{\alpha,\mathrm{deph}} = \frac{1}{3}\text{ },\text{ }p_{\beta,\mathrm{deph}}&=\frac{5}{8}-\frac{1}{8}p \\
    p_{\alpha,\mathrm{ad}}=\frac{1}{3}+\frac{1}{6}p\text{ },\text{ }p_{\beta,\mathrm{ad}}&=\frac{1}{2}+\frac{1}{8}\sqrt{1-p}
\end{aligned}
\end{equation}
From the relations outlined in Eq. \eqref{probability_eq}, one can infer that no matter what noise model carrier $K$ experiences, excluding the case of the dephasing channel where $p_{\alpha,deph}\not\propto p$, protocol $\alpha$ always increases the probability of obtaining the right measurement outcome that localizes maximum entanglement between $A$ and $B$. In contrast, the probability of obtaining the correct measurement outcome to maximize the entanglement in protocol $\beta$ always decreases.

\subsection{Multichannel model}
\label{multi channel model sec}

\begin{figure*}[tp]
    \centering
    $\delta^{\alpha\beta}{\cal N}_{A:B}$\hskip4.5cm$\delta^{\alpha\text{DED}}{\cal N}_{A:B}$\hskip4.5cm$\delta^{\beta\text{DED}}{\cal N}_{A:B}$
    \includegraphics[width=\textwidth]{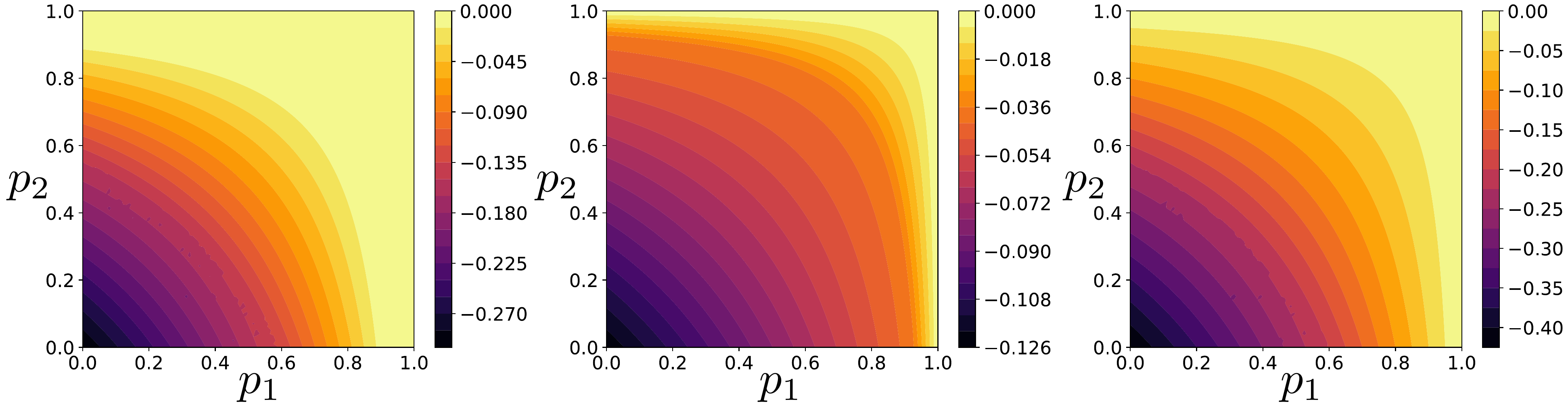}
    \includegraphics[width=\textwidth]{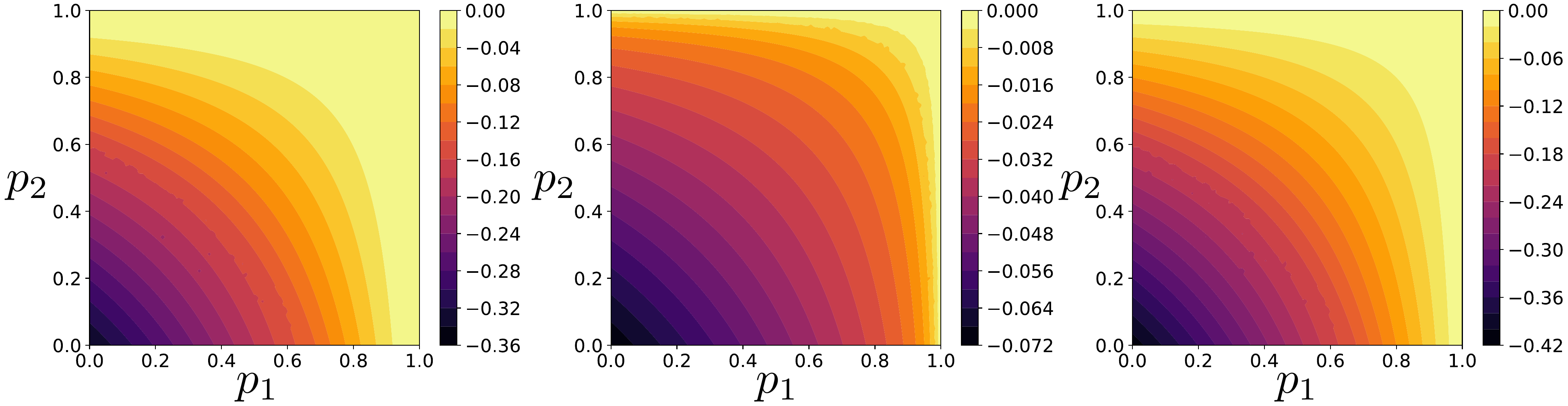}
    \includegraphics[width=\textwidth]{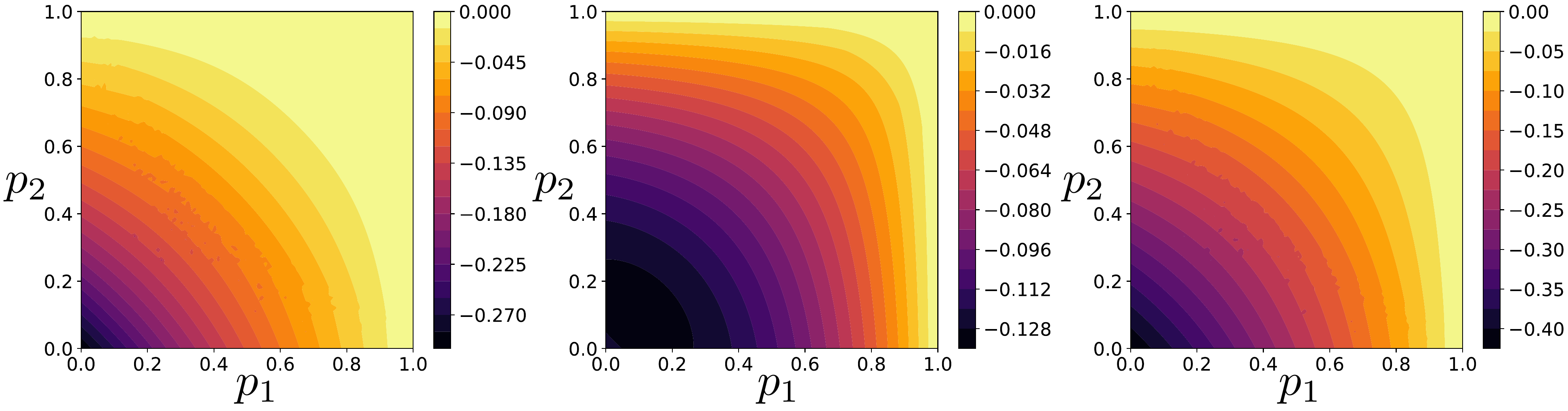}
    \includegraphics[width=\textwidth]{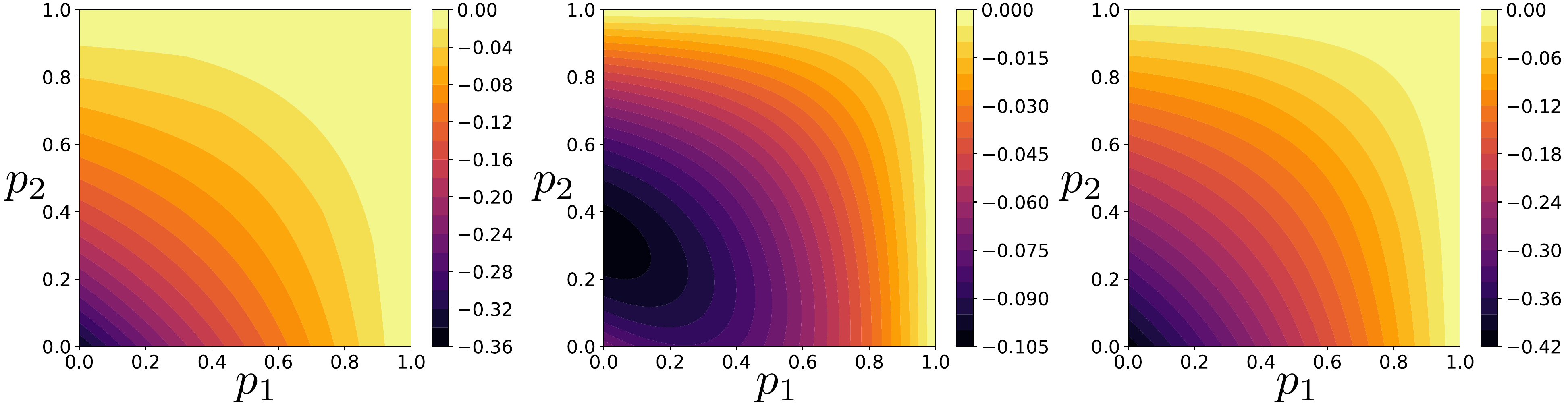}
    \caption{Performance of the DED and EDSS protocols under the effects of noise  with dissimilar channel strengths. We plot the figures of merit $\delta^{ij}{\cal N}_{A:B}$ versus the noise strengths $p_{1,2}$ for a choice of $p_3$ under different noise configurations. 
    First row: $A$ and $B$ are dephased; $K$ experiences depolarizing noise. Second row: $A$ and $B$ are dephased; $K$ is subjected to amplitude damping. Third row: $A$ and $B$ experience amplitude damping; $K$ is depolarized. Fourth row: $A$, $B$, and $K$ all experience amplitude damping noise. All plots are for $p_{1,2} \in [0,1]$ and $p_{3}=0.1$.}
    \label{fig: p3=0.1 contours}
\end{figure*}

\begin{figure*}[tp]
    \centering
    $\delta^{\alpha\beta}{\cal N}_{A:B}$\hskip4.5cm$\delta^{\alpha\text{DED}}{\cal N}_{A:B}$\hskip4.5cm$\delta^{\beta\text{DED}}{\cal N}_{A:B}$
    \includegraphics[width=\textwidth]{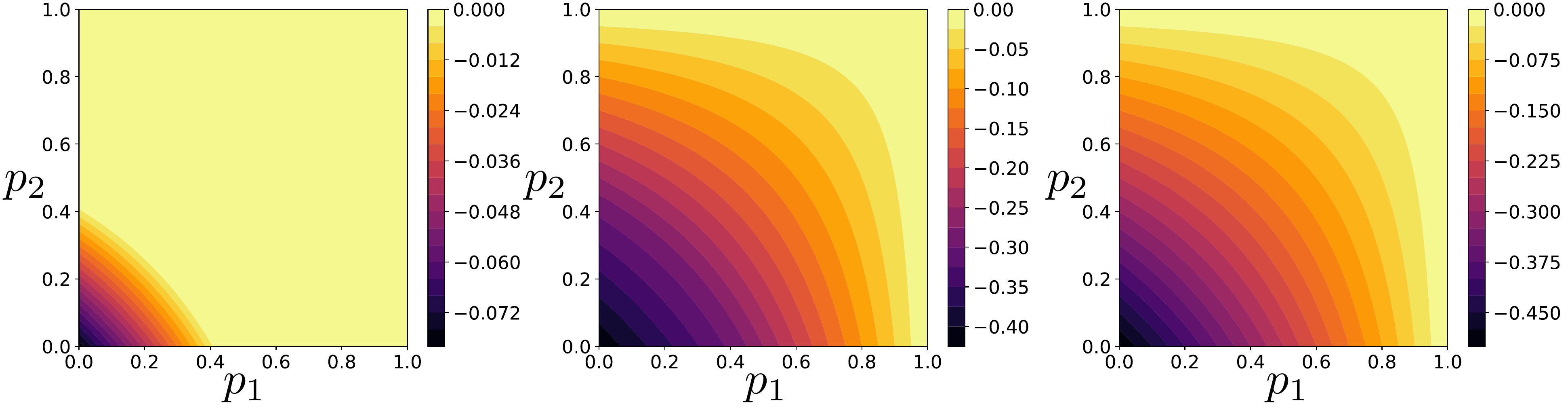}
    \includegraphics[width=\textwidth]{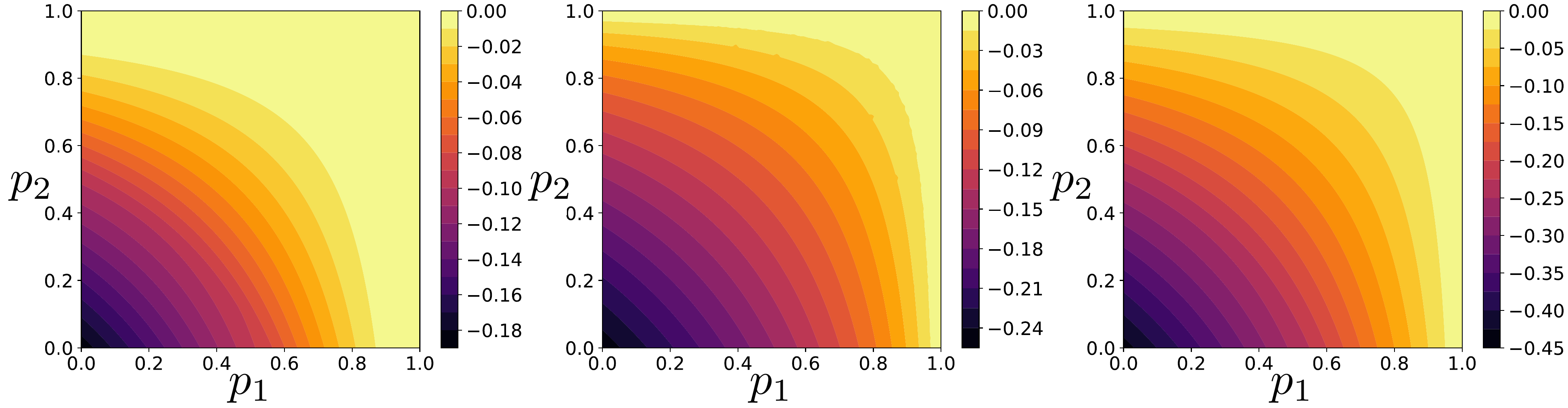}
    \includegraphics[width=\textwidth]{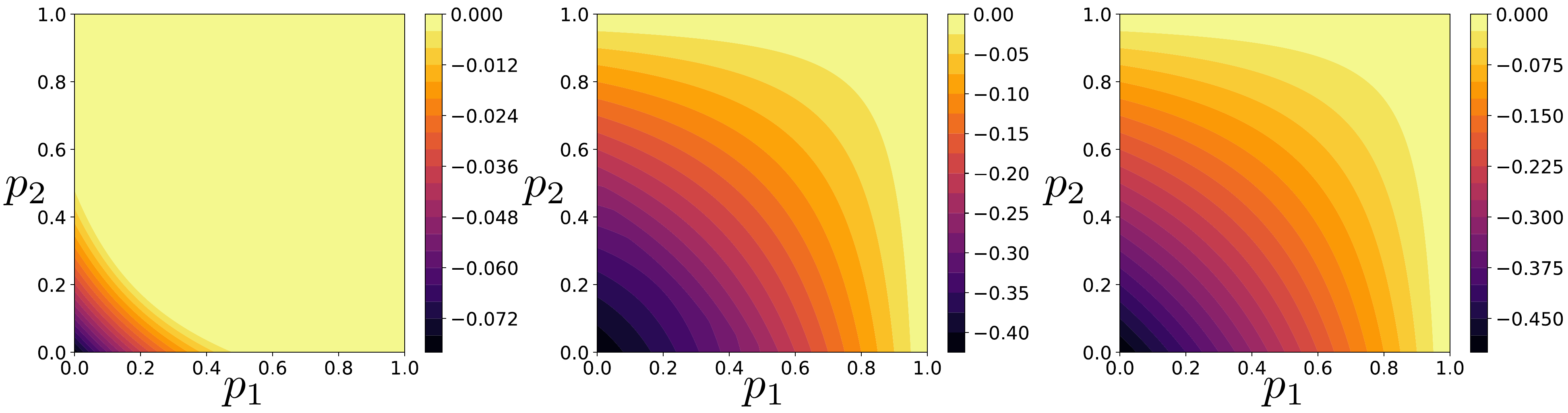}
    \includegraphics[width=\textwidth]{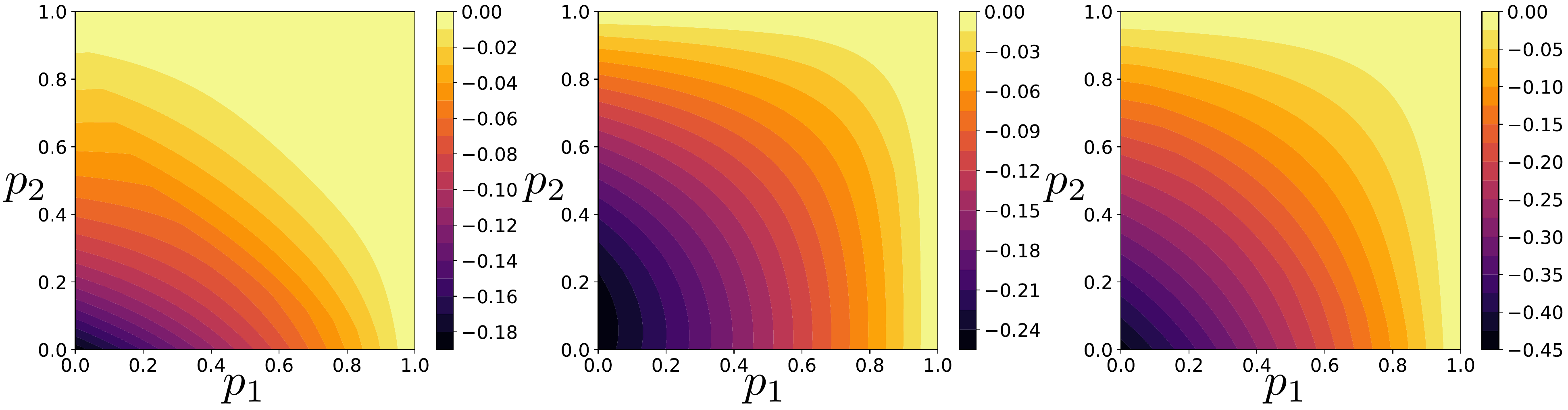}
    \caption{Same as Fig.~\ref{fig: p3=0.1 contours}, but for $p_{3}=0.4$.}
    \label{fig: p3=0.4 contours}
\end{figure*}

Individually, the simple noise models discussed in Sec.~\ref{single channel model sec} are insufficient to realistically simulate the noise experienced by EDSS protocols with the source in the middle architecture outlined in Sec.~\ref{zalm mapping sec}. Whereas in the previous section we considered only noise in the optical channels, we now include the noise to which the memories would be exposed for both DED and EDSS. The qubits encoded in the spin degree of freedom of a memory can experience either dephasing or amplitude damping noise, which would physically stem from mechanisms akin to spontaneous emission from solid-state information carriers. The transmitted separable carrier $K$ can experience either depolarizing or amplitude damping noise, in line with what could be expected from propagating photons across optical fibers. In what follows, we investigate each of the four combinations. Note that, in contrast to Sec.~\ref{single channel model sec}, we now assume that the channels between the ZALM photon source and the memories are noiseless. For all ED schemes, we again assume uncorrelated noise of the same strength $p$, which implies that for the EDSS protocols, the action of the channels on the state of the three-qubit system consisting of qubits $ABK$ reads as defined in Eqs.~\eqref{multi channel eq 1} and \eqref{multi channel eq 2}.

Our results are presented in Fig.~\ref{fig: same_strength_channels plots}, where we analyze up to four iterations of protocol $\beta$. From these plots it is clear that the additional source of noise in the EDSS protocols has a detrimental effect on the entanglement that can be distributed between $A$ and $B$. In each case, since the Bell state emitted by the ZALM source does not undergo depolarizing noise, the DED protocol succeeds in generating entanglement for $p<1$.

Like the analysis conducted in Eq. \eqref{probability_eq}, we studied the probability of obtaining the optimal measurement outcome which maximizes $AB$ entanglement and how this changes with noise when both memories and channel are affected. As one would expect, the success probability increases with noise for protocol $\alpha$ and decreases with noise for protocol $\beta$. Interestingly, if the memory is affected by dephasing noise, then the probability of obtaining the right measurement outcome depends only on the channel noise.

Now we consider the effects of noise models with different strengths to determine when, if at all, the EDSS protocols can outperform DED. We label $p_1$ and $p_2$ the strengths of the noise affecting memories $A$ and $B$ respectively. Furthermore, as the separable carrier in the EDSS protocols is sent through a quantum channel to Bob's site, we assume this channel has strength $p_{3}$. This results in the model
\begin{equation}
    \tilde{\rho}_{ABK}'' = \sum_{k} \big[ \mathbb{I}_{AB} \otimes M_{k , K}(p_{3}) \big] \tilde{\rho}'_{ABK} \big[ \mathbb{I}_{AB} \otimes M^\dag_{k , K}(p_{3}) \big],
\end{equation}
with $\tilde{\rho}'_{ABK}$ resulting from the effect of noise on the $A$-$B$ compound, that is,
\begin{equation}
        \tilde{\rho}'_{ABK} = \sum_{i ,j} \varepsilon''^{ij}_{ABK}(p_{1},p_{2})\,\rho_{ABK}\,\varepsilon''^{ij\dag}_{ABK}(p_{1},p_{2}),
        \end{equation}
with $\varepsilon''^{ij}_{ABK}(p_{1},p_{2}) = M_{i , A}(p_{1}) \otimes M_{j , B}(p_{2}) \otimes \mathbb{I}_{K}$. The DED protocol is now modeled as
\begin{equation}
        \tilde{\rho}_{AB} = \sum_{i ,j} \varepsilon'^{ij}_{AB}(p_{1} , p_{2}) \;|\Phi^{+}\rangle_{AB}\langle\Phi^{+}|\;\varepsilon'^{ij\dag}_{AB}(p_{1} , p_{2}),
        \end{equation}
with $\varepsilon'^{ij}_{AB}(p_{1} , p_{2}) =  M_{i , A}(p_{1}) \otimes M_{j , B}(p_{2})$. For this model, we take $p_{1,2}\in[0 , 1]$, while the value of $p_{3}$ is fixed to either $p_{3}=\{0.1,0.4\}$, as shown in Figs.~\ref{fig: p3=0.1 contours} and ~\ref{fig: p3=0.4 contours}, respectively. This set of values for $p_3$ is chosen merely for illustrative purposes because it allows a clear representation of the results that we have achieved. In this case, only a single iteration of protocol $\beta$ was considered. 

To characterize the performance of the EDSS schemes in the varying noise regimes, the entanglement localized between the cavities is considered for three different cases. The first comparison is the difference between the entanglements localized by protocol $\beta$ and protocol $\alpha$, that is, 
\begin{equation}
\delta^{\alpha\beta}{\cal N}_{A:B}=\mathcal{N}_{A:B}^{\beta}-\mathcal{N}_{A:B}^{\alpha}.
\end{equation}
As protocol $\beta$ will localize less entanglement,  we gauge its performance with respect to protocol $\alpha$. The other comparisons we perform involve the differences between the entanglement localized by protocol $j=\alpha,\beta$ and DED, namely,
\begin{equation}
\label{jDED}
\delta^{jDED}{\cal N}_{A:B}=\min[0,\mathcal{N}_{A:B}^{j}-\mathcal{N}_{A:B}^{DED}].
\end{equation}
The choice in Eq.~\eqref{jDED} to consider the smallest value between zero and the difference in entanglement established by pairs of protocols is made to emphasize the regions in the parameter space of noise strengths where a DED approach is more advantageous than either of the two EDSS schemes that we have addressed. 

Figure~\ref{fig: p3=0.1 contours} shows that dephasing the discordant state of the memories results in protocol $\alpha$ being more advantageous than $\beta$ for lower values of $p_{1}$ and $p_{2}$, as seen by the larger region of dark color. The analysis of the relative performance between DED and the EDSS-based protocols $\alpha$ and $\beta$ reveals that the $\alpha$ strategy delivers values of entanglement only marginally different from those of DED, while $\beta$ is inferior in performance compared with the latter. Moreover, as protocol $\beta$ can localize only up to $N_{A:B}^{\beta}=0.1$ entanglement between $A$ and $B$, we infer that this scheme is virtually ineffective for much of the range of values that $p_{1}$ and $p_{2}$ can take. This is evidenced by the fact that the plots in the right column of Fig.~\ref{fig: p3=0.1 contours} are very similar to each other. 

In Fig.~\ref{fig: p3=0.4 contours}, which refers to the choice of $p_3=0.4$, one can see that subjecting the separable carrier $K$ to depolarizing noise reduces the discordant advantage of both EDSS protocols. Instead, amplitude damping allows both schemes to localize entanglement in regions with higher values of $p_{1}$ and $p_{2}$. This hearkens back to the robustness of quantum discord as a resource, as presented in Fig.~\ref{fig:channels_figure} \textbf{(c)}, against this sort of noise. This distinctive feature carries over into the comparison between DED and protocol $\alpha$ for high values of $p_{1}$ and $p_{2}$ when the carrier is exposed to amplitude damping, whereas depolarizing $K$ results in a larger difference between the entanglement that is achieved by the two schemes. As before, $\delta^{\beta DED}{\cal N}_{A:B}$ appears to be largely insensitive to the channel configuration.  

The overall message brought forward by this analysis is that the ZALM-like architecture cannot outperform DED and fails to utilize the unique advantage of quantum discord's robustness in noisy environments.

\section{Conclusions}
\label{conc}

We showed that ZALM-based architectures can be extended to methods of entanglement distribution beyond directly sending a Bell state. The methods used to transfer maximally entangled states from photon polarization to spin degrees of freedom can equally be used to map any state of choice. Therefore, we can, in fact, adapt a ZALM framework to carry out EDSS protocols, which require only transmitting discordant states to the memories and sending a separable carrier photon between them.

We then analyzed the advantages and limitations of using EDSS within a ZALM setup. On exposing the key resources of the various protocols to different types of noise, we found that EDSS consistently outperformed the original Bell-state distribution protocol. EDSS protocol $\alpha$, which requires initial classical correlations between the memories and the carrier, can generate the most entanglement in the case of depolarizing or dephasing noise. Weak amplitude damping still favors protocol $\alpha$ while protocol $\beta$, in which the carrier and memories initially share zero correlations, is the best choice under strong amplitude damping conditions.

However, such benefits depend heavily on the noise model employed, as we discovered when taking a different approach that explicitly focuses on the noise affecting the memories. The necessary extra step of sending the carrier through a noisy channel exposes the EDSS schemes to additional noise compared to the direct one. Under these assumptions, thus, DED is the optimal method irrespective of the noise being considered.

We conclude that, due to their ability to adapt to varied protocols, ZALM architectures provide promising avenues when considering the overarching goal of developing a quantum internet. Further investigations into their use in quantum networks are thus necessary. A potential pathway could be to investigate whether we can create a responsive ZALM setup which detects the environmental conditions of the network and the resources at hand in order to carry out the optimal entanglement distribution protocol.

The data generated as part of this work are available from Ref.~\cite{GitHub}  and, upon reasonable requests, from the authors.

\acknowledgements
C.J.C., A.G.H. and M.P. are grateful to the members of the Quantum Theory Group at the Department of Physics and Chemistry of the University of  Palermo for comments and insight. We acknowledge support from the European Union's Horizon Europe EIC-Pathfinder project QuCoM (Grant No. 101046973), the Royal Society Wolfson Fellowship (Grant No. RSWF/R3/183013), the United Kingdom's EPSRC (Grant No. EP/T028424/1), and the Department for the Economy Northern Ireland under the US-Ireland R\&D Partnership Programme. 

\bibliography{refs.bib}

\end{document}